\documentclass[aps,pre,groupedaddress,showpacs,showkeys,twocolumn]{revtex4}
\usepackage{eepic}
\usepackage[]{natbib}
\usepackage{graphicx}
\usepackage{dcolumn}
\usepackage{amsmath}
\usepackage[latin1]{inputenc}

\begin{document}

\title[]{Escaping from cycles through a glass transition}

\author{\firstname{Sebastian} \surname{Risau-Gusman}}
\email{srisau@if.ufrgs.br}
\affiliation{Instituto de F\'{\i}sica,
Universidade Federal do Rio Grande do Sul, 91501-970, Porto Alegre, RS, Brazil.}

\author{\firstname{Alexandre} S. \surname{Martinez}}
\email{amartinez@ffclrp.usp.br}
\homepage{http://www.fisicamedica.com.br/martinez/}

\author{\firstname{Osame} \surname{Kinouchi}}
\email{osame@dfm.ffclrp.usp.br}

\affiliation{Faculdade de Filosofia, Ci\^encias e Letras de Ribeir\~ao Preto,
Universidade de S\~ao Paulo \\ Av. Bandeirantes 3900, 14040-901,
Ribeir\~ao Preto, SP, Brazil.}

\begin{abstract}
A random walk is performed over a disordered media composed of $N$
sites random and uniformly distributed inside a $d$-dimensional
hypercube. The walker cannot remain in the same site and hops to
one of its $n$ neighboring sites with a transition probability
that depends on the distance $D$ between sites according to a cost
function $E(D)$. The stochasticity level is parameterized by a
formal temperature $T$. In the case $T = 0$, the walk is
deterministic and ergodicity is broken: the phase space is divided
in a ${\cal O}(N)$ number of attractor basins of 2-cycles that
trap the walker. For $d = 1$, analytic results indicate the
existence of a glass  transition at $T_1 = 1/2$ as $N \rightarrow
\infty$. Below $T_1$, the average trapping time in 2-cycles
diverges and out-of-equilibrium behavior appears. Similar glass
transitions occur in higher dimensions when the right cost
function is chosen. We also present some results for the
statistics of distances for Poisson  spatial point processes.

\end{abstract}

\keywords{random walks,
          trap model,
          exploratory behavior,
          rugged landscapes,
          glass transition,
          disordered media.
         }

\pacs{05.40.Fb,  
      02.50.-r   
      05.90.+m,  
      87.15.Aa  
     }

\maketitle

\section{Introduction}

Random walks in random media constitute an interesting
mathematical problem with several physical applications
\cite{fisher:1984, barkema:2001}.
The study of deterministic walks in random media is also a fascinating topic but presents difficulties (nonergodicity, for instance) common to the area of dynamical systems.
The statistics of such deterministic walks, which present transient and cycling parts, have been much less investigated \cite{grassberger:92, gale:95, lima_prl2001, kinouchi_phys_a} up to date.
Here we investigate the transition from deterministic to random walks in a disordered medium.
We find that escaping from a cycle dominated phase is done through a glass transition in the ``weak ergodicity breaking'' scenario present in Limoge-Bocquet \cite{limoge} and Bouchaud's trap models \cite{dyre:1987,bouchaud:1992,dyre:1995,monthus:1996,bouchaud:2001,bouchaud:2002}.

Any spin lattice model with deterministic parallel update may be viewed as performing a deterministic walk in its phase space.
Examples are the Little model \cite{little, fontanari} (a ``Hopfield neural network''
with parallel dynamics), asymmetric neural networks with parallel update \cite{bastolla:1998},  Kauffmann Boolean networks \cite{bastolla:1996}, etc.
Since such systems have a huge and complex phase space, it is interesting to compare
their behavior to other simpler dynamical systems with quenched disorder.
This is the rationale for studying toy models as the Random Map~\cite{derrida} that shares some generic statistical and dynamical properties with those more complex systems.

A phase composed only by 2-cycles (i.e. cycles involving only two states) occurs in the neural and Kauffman networks cited above.
However, such 2-cycle phase is not present in the Random Map model.
Recently, we introduced a simple dynamical system with quenched disorder that presents such a phase: the {\em Tourist Walk} \cite{lima_prl2001,stanley_2001, kinouchi_phys_a}.
It is a deterministic walk over a set of $N$ points.
The points may be sites in a $d$-dimensional Euclidean space \cite{lima_prl2001, kinouchi_phys_a} or nodes in a graph with ordered neighborhood (for example, a thesaurus graph \cite{kinouchi_phys_a}).
The {\em Tourist Rule} is: at each time step, the walker must go to the nearest site not visited in the past $\tau$ steps (a self-avoiding window).
We have found that, for any $\tau$ window, the phase space is divided in a ${\cal O}(N)$ number of basins of $p$-cycles attractors. In particular, when $\tau = 0$, all attractors
have period $p=2$.
For $\tau =1$, in the $d \rightarrow \infty$ limit, the Tourist Walk presents a power law distribution of cycle periods of the form $P(p) \propto p^{-1}$, with a cutoff which scales with
$\sqrt{N}$. This is similar to the distribution present in the Random Map model~\cite{derrida}.

The above lattice systems also have been studied with stochastic
dynamics. We expect that above a certain temperature the dynamics
is no longer dominated by the cycles. It is not clear how to
introduce a stochastic dynamics in the Random Map but we found
that it is easy (and even natural) to introduce a formal
temperature $T$ in the Tourist Walk such that the deterministic
limit is recovered when $T = 0$. The main question we address here
is how ergodicity could be recovered from such deterministic
dynamical systems with quenched disorder. The introduction of
dynamical stochasticity is necessary to break the cycles, but a
low level of stochasticity may not be sufficient for full
ergodicity recovery. Indeed, we have found a glass-like transition
at a finite stochasticity level (temperature) $T_c$. We expect
that a similar behavior is present in, for example, stochastic
versions of Kauffman networks.

Here we consider a stochastic Tourist Walk where the walker, at
site $i$, is allowed to jump to its first $n$ neighbors with a
transition probability proportional to $\exp[- E(D_{ij})/T]$. The
stochasticity level is parameterized by $T$ (a formal
temperature) and the cost function $E(D_{ij})$ is a monotone
increasing function of the distance between the sites. When the
formal temperature $T$ is zero, one has a simple Tourist Walk
without self-avoiding window ($\tau = 0$): the walker always goes
to the nearest neighboring site, eventually being trapped in a
2-cycle when a pair of reciprocally nearest neighbors is found.
For $T > 0$, the 2-cycles are no longer stable, being
characterized by a distribution of trapping times $P(t_r)$.
However, for low values of $T$, the walker is in an
out-of-equilibrium regime,  presenting a divergent average
trapping time when $N\rightarrow \infty$ (``weak ergodicity
breaking'' scenario
\cite{limoge,dyre:1987,bouchaud:1992,dyre:1995,monthus:1996,bouchaud:2001,bouchaud:2002}).
Increasing $T$, one is able to study the ergodicity recovery
transition where the average trapping time becomes finite. For
one-dimensional systems this occurs at $T_1 = 1/2$ for all values
of the connectivity $n$.

The paper is organized as follows. In Sec.~\ref{sec:desc} we
define the model and set the notation. In Sec.~\ref{sec:geo_net}
we show that the network structure, where hops occur, forms a
directed graph. This graph presents interesting sets of points
that we call {\em sinks} and {\em sources}. We present analytic
results on the density of sinks of a given size. In
Sec.~\ref{sec:stat} we obtain some joint distributions of
distances for Poisson spatial point process. These results are
used in the calculation of bounds for the average trapping time in
2-cycles. In Sec.~\ref{sec:phase_trans} we show that in
one-dimensional systems there is a glass transition at $T_1 =
1/2$. Below $T_1$, the average trapping time diverges and the system
falls into an out-of-equilibrium state. A discussion on the
existence of such glass transitions in higher dimensions is presented.
Exploratory behavior with a power-law hopping process is discussed
via the introduction of a proper cost function. Our final remarks
and conclusions are offered in Sec.~\ref{sec:conc}.

\section{Description of the model}
\label{sec:desc}

The model is defined by an underlying disordered lattice composed of $N$ sites, whose coordinates $x_i^{(k)}$ ($i=1,\ldots, N$; $k = 1,\ldots, d$) are distributed uniformly in the $d$-dimensional unitary volume $[0,1]^d$.
From the sites coordinates, one constructs the Euclidean distance matrix:
\begin{equation}
D_{ij} \equiv N^{1/d}  \left\{ \sum_{k=1}^d \left[ x_i^{(k)}-x_j^{(k)}
\right]^2 \right\}^{1/2} \; ,
\end{equation}
where the factor $N^{1/d}$ has been introduced to simulate a constant density of points, when the thermodynamical limit $N \rightarrow \infty$ is taken. In this limit, the point distribution is equivalent to that of a Poisson process in $\Re^d$, using the geometrical distance.

Now consider a hopping process between sites.
The transition probability from site $i$ to site $j$ is defined as:
\begin{equation}
\label{eq:weight}
W_{i \rightarrow j}(\beta)  =  \frac{e^{ -\beta E(D_{ij})}}{Z_i} \; ,
\end{equation}
where the normalization factor is:
\begin{equation}
Z_i(\beta)  =  \sum_{j \in nn} e^{-\beta E(D_{ij})} \; \label{Z} ,
\end{equation}
with the sum running over the $n$ nearest neighbors ($nn$) of point $i$.
The cost function $E(D_{ij})$ depends only on the distances $D_{ij}$, and is a monotone increasing function.

Several scenarios can be envisaged.
If $E(D_{ij}) = D_{ij}$ then the transition probability decays exponentially with the distance.
If $E(D_{ij})= D_{ij}^2$ a gaussian decay is obtained for $W_{i \rightarrow j}$.
If $E(D_{ij}) = \ln {D_{ij}}$, then $W_{i \rightarrow j} \propto D_{ij}^{- \beta}$ decays as a power law of the distance.

Notice that the transition probabilities are not symmetric ($W_{i \rightarrow j} \neq W_{j \rightarrow i}$) because the normalization factors are different ($Z_i \neq Z_j$): the neighbors of $i$ are not the same as the neighbors of $j$.
Furthermore, the walker is not allowed to remain at the same site,
i.e., $W_{i \rightarrow i}=0$ (which is equivalent to setting $D_{ii} = \infty$ in the distance matrix).

The stochastic parameter or ``temperature'' $T = 1/\beta$ regulates the distance dependent bias of the model.
When $\beta \rightarrow 0$, all the transition probabilities are equal and one obtains unbiased hopping inside a disordered lattice.
When $\beta \rightarrow \infty$, in Eq.~(\ref{Z}) the term corresponding to the first nearest neighbor dominates, yielding either $W_{i \rightarrow j}= 1$, if the site $j$ is the nearest neighbor of site $i$, or $W_{i \rightarrow j}= 0$, otherwise.
The tourist goes imperatively to the nearest site, relative to the actual position, thus performing a deterministic walk~\cite{lima_prl2001} without a self-avoiding window (that is, with $\tau=0$).

Observe that the model has not been defined in a regular lattice because in that case the number of nearest neighbors is degenerated (greater than one) and determinism cannot be recovered.
The prohibition for the walker to remain at the same site is needed to fulfill the Tourist Walk rule.

When $n=N-1$, any site is accessible from any other site.
However, if $n$ is small, the underlying network is a diluted directed graph.
Such graphs have a complicated topology: some sites may not be accessible from other sites and the graph may be disconnected.
Thus, we will first study the topology induced in this graph by a finite degree $n$ of outgoing links.
Afterwards, we present the dynamical (``glass'') transition that occurs in the walks performed on this graph, when $T$ is sufficiently low.

\section{Topology of the underlying network}
\label{sec:geo_net}

We have assumed that transitions occur only to the first $n$ nearest neighbors of each site $i$.
Viewing each transition term $W_{i \rightarrow j}$ as the weight of the link between sites $i$ and $j$, we obtain a directed network or graph.
The degree distribution of outgoing links is a delta function, $P(k_{out})=\delta_{k,n}$, and the distribution of ingoing links $P(k_{in})$ is binomial.
Notice however that this is not a random graph, as it is restricted to a $d$-dimensional Euclidean space, and thus it cannot be fully characterized by using only the distribution of ingoing and outgoing links.
This graph is the underlying network where deterministic Tourist Walks occurs (if we set $n=\tau+1$), so it has been called a {\em Tourist Graph}.

\subsection{Sinks and Sources}

For all finite values of $n$, topological constraints induce special sets of points.
For example, using $n=2$ and $d=2$, consider a set of three points such that each point has the other two points as nearest neighbors.
All the outgoing links of such points remain inside this set of three points, forming a triangle.
If this triangle does not receive any ingoing link from outside points, this triangle is an isolated cluster.
If it receives some ingoing links, it acts as an absolute trap: the cluster may be attained from outside, but there is no possibility to escape from it, since there is no transition terms $W_{i \rightarrow j}$ to any site $j$ outside the triangle.
Such sets, that do not present outgoing links, we call {\em sinks}.
When these sinks are {\em minimal} (i.e. there are no smaller sinks inside them), they correspond to what are called {\em irreducible closed sets} in the terminology of Markov processes \cite{feller}.
Sinks are composed of persistent states and the rest of the graph is composed of transient ones ~\cite{feller}.
Notice that, since there are always $n$ outgoing links from each point, a single site cannot be a sink.
Sinks have a minimal size of $n+1$ points.

Consider also a set of points (or even a single point) that present outgoing links to points outside the given set, but with no ingoing links from exterior points.
Such sets are called {\em sources}.
A walk may be started inside a source, but if the walker exits this set, she cannot return to it any longer.

\begin{figure}[htb]
\begin{center}
\includegraphics[angle=-90,width=8.5cm]{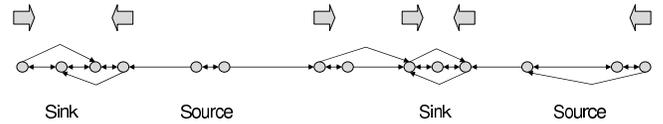}
\end{center}
\caption{Sinks and sources in one dimension. Large arrows represent walls.}
\label{fig:1}
\end{figure}

Sinks and sources are easily visualized in dimension $d = 1$ (see Fig. \ref{fig:1}) because we can define special points (``walls'').
A site $w$ is a (semi-permeable) {\em wall} whenever all of its $n$ outgoing links point along the same direction.
However, $w$ may have ingoing links from the opposite direction, that is, the walls behave as semi-permeable membranes: they allow the flux of walkers from one side to another but not the other way around.
Thus the walls are characterized by the directions in which the walker flux is allowed.
This is indicated in Fig. \ref{fig:1} by arrows and, as a shorthand, we say that the wall ``points'' to the direction indicated by the large arrows.

In $d=1$ systems, two consecutive walls define different regions on the disordered lattice depending on their orientation.
If the two walls point to the center of a region (inwards) ($\mid \Rightarrow \Leftarrow \mid$), the set of points comprised by all the points between the walls is a minimal sink.
Once a walker attains a sink, there is no escape: the only permissible movement is to  wander stochastically inside it.
If the two consecutive walls point outwards a region ($\Leftarrow \mid \mid \Rightarrow$), this region is a source. Once a walker escapes from such a region, it cannot return.

Going to a higher level of description, our graph can be characterized by the sets discussed in Refs.~\onlinecite{dorogov:2001} and~\onlinecite{newman:2001}.
For $d>1$ and large enough values of $n$ the graph has a {\em giant weakly connected component\/} (GWCC) made of points connected by links (without considering their direction).
Inside this set, there is a {\em giant strongly connected component\/} (GSCC), where every point is accessible from every other point of the set through a directed path.
This is the set where, most probably, a random walker roams until she eventually falls into a sink.
By definition~\cite{newman:2001}, the GSCC does not contain either sinks or sources.
The (disjoint) set of sources whose outgoing links lead to the GSCC is the  {\em in-component} (IN), i. e., the set of points from where the GSCC is reachable but cannot be reached from it.
If the walker is initially placed in the IN, she will leave this set for the GSCC in a relatively small amount of time.
The disjoint set of sinks that can be reached from the GSCC is called the {\em out-component} (OUT).
This is the set where almost every random walker will end up.
These are the greatest sets in our graph, but other smaller sets are possible.
For example, the set of all the sources that are not connected to the GSCC, but to other sinks.
Or the set of sinks that cannot be reached from any other point in the graph.

In $d=1$, due to the very special topology of $\Re$, things are different.
Sinks disconnect the graph (nothing passes from one side to the other of a sink) and a GSCC cannot be formed.

Notice that the presence of sources and sinks occurs even in the high temperature limit $(T \rightarrow \infty$).
It is a topological property, not a dynamical one, emerging from the fact that each site has only $n$ outgoing links.
The glass transition discussed in Section \ref{sec:phase_trans} has nothing to do with such topological constraints.
Simulations show that it is present even in the fully connected case $n=N-1$.
Sinks are an undesirable (although inevitable) factor that complicates the study of the glass transition in our random walk. On the other hand, finite connectivity allows us to extract some analytical results, which is not the case for $n \rightarrow \infty$.

\subsection{Density of sinks}
\label{sec:backbone}

Next we calculate the density of sinks in one dimension and ask for the conditions which enable the emergence of a single macroscopic connected region where any point is
reachable from any other.

In $d=1$, a sink is formed by two inward walls where each wall site has $n$ outgoing links.
Thus, the smallest sink comprises $n+1$ points.
It is possible to calculate explicitly the density (also called concentration) $C_m(n)$ of sinks with sizes $n+m$ for $m$ small.

Consider a set of $n+3$ consecutive points, relabeled as $i=1,\ldots,n+3$ and with coordinates $x_i$.
Let $X_i = x_{i+1} - x_i$ be the distance separating two consecutive points.
We first calculate the probability that point $i=2$ is a wall pointing to the right and that, simultaneously, point $i=n+2$ is a wall pointing to the left with no wall between these two walls.
If $X_1$ is greater than the total distance between points $i=2$ and $i=n+2$, then point $i=2$ is a left wall pointing to the right.
Similarly, if $X_{n+2}$ is greater than the total distance between points $i=2$ and $i=n+2$, then point $i=n+2$ is a right wall pointing to the left.
Since points are randomly spaced along the line (Poisson process), the probability of having a distance $X_i$ is: $P(X_i)=e^{-X_i}$.
Thus, the concentration of sinks having $n+1$ points is:
\begin{eqnarray}
\nonumber
C_1(n) & = &  \int \prod_{i=1}^{n+2} (\mbox{d} X_i \, e^{-X_i})  \Theta(X_1-\sum_{i=2}^{n+1} X_i) \\
       &   & \times \Theta(X_{n+2}-\sum_{i=2}^{n+1} X_i)  =  \frac{1}{3^{n}} \; .
\label{eq:C1}
\end{eqnarray}

Similar (but more cumbersome) calculations give the concentration for
the next two sink sizes ( $n+2$ and $n+3$).
For $n \geq 2$:
\begin{eqnarray}
C_2(n) &=& \frac{1}{4} \left(\frac{1}{3^{n-1}}- \frac{1}{5^{n-1}} \right) \\
C_3(n) &=& \frac{1}{16}\left(\frac{1}{3^{n-2}}- \frac{1}{5^{n-2}}
+\frac{1}{9}\,\frac{1}{7^{n-2}}+\frac{1}{12} \, \frac{1}{9^{n-2}} \right) \:.
\end{eqnarray}

For large values of $n$, $C_2(n)/C_1(n) \rightarrow 3/4$
and $C_3(n)/C_2(n) \rightarrow 3/4$. One may conjecture that, asymptotically, this will hold true for all size
concentrations. In other words, $C_{m+1}(n)/C_m(n) \rightarrow 3/4$ for all values of $m$.
If this conjecture is valid, the fraction of
points that belong to sinks is, asymptotically,
$F(n) = \sum_{m=1}^{\infty} (m + n) C_m(n) = (4n+20)/3^n$ and the total concentration of sinks is $C(n) = \sum_{m=1}^{\infty} C_m(n) = 4/3^n$.
This means that for large values of $n$ the graph is effectively disconnected and the clusters have on average $[1 - F(n)]/C(n) = (3^n-4n-20)/4$ points.

This result can be used to estimate
the minimum value of $n$ for which there exists a single cluster of
size ${\cal O}(N)$.
If one imposes $N C(n) = 1$ one gets $n_{min} = \ln(4N)/ \ln 3$.
This implies that it is not necessary to take $n = N-1$ to have a single
macroscopic cluster: it can be obtained with very high probability for low values [${\cal O} (\ln N)$]
of $n$.

It is easy to see that for higher dimensions and finite $n$ there will also be a finite fraction of sinks and sources, in the thermodynamical limit $N \rightarrow \infty$. However, numerical evidence presented in Sec.~\ref{sec:app1} seems to imply that the glass transition studied in Sec.\ref{sec:phase_trans} occurs in the GSCC, that is, outside the sinks and sources.

\section{Couple Statistics}
\label{sec:stat}

In this section we obtain the density of couples (which form 2-cycles traps)
in $d$ dimensions and we derive probability density functions for distances inside couples and between couples and their neighbors.

\subsection{Density of couples}

With $T = 0$, starting the walk at a given site, after a fast transient \cite{idiart:2002}, the walker eventually reaches two ``reflexive nearest neighbors'' (or ``couple'', for short) \cite{cox}.
A couple $a \leftrightarrow b$ is formed when site $a$ is the nearest neighbor of site $b$ and $b$ is the nearest neighbor of $a$.
Once a couple is attained, the walker is trapped in a 2-cycle attractor ($a-b-a-b-\ldots$).
As shown in Ref. \onlinecite{kinouchi_phys_a}, for $d=1$, the expected number of couples is $M(1) = N/3$, and for general values of $d$ one has the couple density:
\begin{equation}
\frac{M(d)}{N} = \frac{1}{2} P_d(a \leftrightarrow b)  =  \frac{1}{2[1+p_d(1)]} \;,
\label{eq:couple_density}
\end{equation}
where $P_d(a \leftrightarrow b)$ is the probability that a point belongs to a couple and
\begin{equation}
\label{eq:pd(x)}
p_d(x) = I_{\left(\frac{x}{2}\right)^2}\left( \frac{1}{2}, \frac{d+1}{2} \right) \; ,
\end{equation}
with the definitions~\cite{stegun}:
\begin{eqnarray}
\label{eq:incompletebeta}
I_z(a,b)  & = & \frac{1}{B(a,b)} \; \int_{0}^{z} \mbox{d}t \; t^{a-1}(1-t)^{b-1} \; , \\
\label{eq:beta}
B(a,b)  & = & \int_{0}^{1} \mbox{d}t \; t^{a-1}(1-t)^{b-1}
          =   \frac{\Gamma(a)\Gamma(b)}{\Gamma(a+b)} \; , \\
\label{eq:gamma}
\Gamma(z) & = & \int_0^{\infty} \mbox{d}t \; t^{z-1} e^{-t} \; ,
\end{eqnarray}
where $I_z(a,b)$ is the incomplete beta function, $B(a,b)$
is the beta function and $\Gamma(z)$ is the gamma function.
The function $p_d(x)$ is given in Table \ref{tab:1} for some values of $d$,  and its
geometrical meaning is presented below (Sec. \ref{sec:cddf}).

\begin{table*}[ht]
\begin{tabular}{c|c|c| c}
\hline
$d$      &  $p_d(x)$   & $p_d(1)$ & $\partial V_{\cup}(D_1,D_0)/\partial D_1$ \\
\hline
$1$      &  $\frac{x}{2}$  & $\frac{1}{2}$ & $2$ \\
$2$      &  $\frac{2}{\pi} \left[ \frac{x}{2} \sqrt{1 - \left( \frac{x}{2} \right)^2} + \arctan \sqrt{\frac{(x/2)^2}{1 - (x/2)^2}} \right] $ & $\frac{3^{3/2} + 2 \pi}{6 \pi}$ & $2 \pi D_1  + 4 D_1 \arctan\sqrt{\frac{[D_1/(2D_0)]^2}{1 - [D_1/(2D_0)]^2}}$ \\
$3$      &  $\frac{3}{2} \left[ \frac{x}{2} - \frac{1}{3} \left(\frac{x}{2} \right)^3 \right] $                                              & $\frac{11}{16}$ & $2 \pi D_1 [2 D_1 + D_0]$ \\
$\vdots$ &  $\vdots$ & $\vdots$ & \vdots\\
\hline
\end{tabular}
\caption{Expression of function $p_d(x)$ for some values of the dimension $d$ (see Eq.~(\ref{eq:pd(x)})) This generalizes Table 2 of Ref. \onlinecite{kinouchi_phys_a}. For the last column see Eq.~(\ref{eq:der_Vu}).}
\label{tab:1}
\end{table*}

The couples trap permanently the walker only at $T=0$.
Notice that, in principle, there is no relationship between couples and sinks (for example, couples may be sources).
However, the special case $n=1$ (a single outgoing link for each point) is equivalent to a $T=0$
scenario. In this case, sinks and couples are identical.
From Eq.~(\ref{eq:C1}), $C_1(1)=1/3$ gives the concentration of
couples, which is compatible with Eq.~(\ref{eq:couple_density}), for $d=1$.

As we are using a spatial Poisson process, the probability of a
point being located in a given volume is proportional to the
volume itself and there is no correlation between points. Next we
present some analytical results on the distance statistics of
point processes in $\Re^d$ that could be of general interest (for
example, in studies of the Travelling Salesman Problem
\cite{Percus}).

\subsection{Couple distance distribution function}
\label{sec:cddf}
First we calculate the probability density function (p.d.f.) for the distances between the points of a couple.
Following Cox \cite{cox}, without loss of generality,
we consider a unitary intensity Poisson process in $d$ dimensions.
For such a process the probability of having no points inside a hypersphere of volume $V$ and radius $D$ is $e^{-V}$.
Then, the p.d.f. of finding the nearest neighbor of a particular site $a$ at a distance $D$ is:
\begin{equation}
P_{nn}(D)  =  \frac{\mbox{d} V(D)}{\mbox{d} D} \; e^{-V(D)} =  A_d d D^{d-1} \;  e^{-V(D)}  \; ,
\label{Pnna}
\end{equation}
with
\begin{eqnarray}
\nonumber
V(D) & = & A_d \; D^d  \\
A_d  & = & \frac{\pi^{d/2}}{\Gamma(d/2+1)} \; .
\label{eq:ad}
\end{eqnarray}

Consider also the volume of the union of two $d$-spheres of radius $D$
centered at sites $a$ and $b$ separated by a distance $D_0$  (see Fig.~\ref{fig:2}):
\begin{equation}
V_{\cup}(D,D_0)   =  2V(D) - V_{\cap}(D,D_0) \:,
\end{equation}
where $V_{\cap}(D,D_0)$ is the volume of the intersection of two $d$-spheres whose centers are separated by a distance $D_0$.
Using the variable
$x= \cos \phi$ (see Fig.~\ref{fig:2}), we get:
\begin{eqnarray}
\nonumber
V_{\cap}(D,D_0) & = & \frac{2 V(D)}{B \left(\frac{1}{2},\frac{d+1}{2} \right)}
\; \int_{\frac{D_0}{2D}}^{1} \mbox{d}x \; (1-x^2)^{\frac{d-1}{2}} \nonumber \\
              & = & \frac{  V(D)}{B \left(\frac{1}{2},\frac{d+1}{2} \right)}
\; \int_{\left( \frac{D_0}{2D} \right)^2}^{1} \mbox{d}t \; t^{- \frac{1}{2}}(1-t)^{\frac{d-1}{2}} \; ,
\nonumber\\
& = &  V(D) \left[ 1 -  p_d\left(\frac{D_0}{D} \right) \right] \:.
\end{eqnarray}
Notice that $p_d(x) = [V(D)-V_{\cap}(D,D_0)]/V(D)$, is the volume fraction occupied by the ``hyper-crescent''.
Thus, the total volume is:
\begin{equation}
V_{\cup}(D,D_0)   =  V(D) \left[ 1 +  p_d\left(\frac{D_0}{D} \right) \right] \; .
\end{equation}

\begin{center}
\unitlength = 0.5 mm
\begin{figure}[htb]
\begin{picture}(150,120)
\put( 50,60){\arc{120}{ 1.15}{ -1.15}}
\put(100,60){\arc{120}{-2.}{  2.}}
\put( 50,60){\blacken\circle{3}}
\put( 50,60){\blacken\circle{2.5}}
\put( 50,60){\blacken\circle{2}}
\put( 50,60){\blacken\circle{1.5}}
\put( 50,60){\blacken\circle{1}}
\put( 50,60){\circle{100}}
\put( 50,60){\vector(1,1){40}}
\put( 75,95){$D$}
\put(100,60){\blacken\circle{3}}
\put(100,60){\blacken\circle{2.5}}
\put(100,60){\blacken\circle{2}}
\put(100,60){\blacken\circle{1.5}}
\put(100,60){\blacken\circle{1}}
\put(100,60){\circle{100}}
\put( 50,60){\vector(1,0){50}}
\put(100,60){\vector(-1,0){50}}
\put( 60,65){$\phi$}
\put( 75,65){$D_0$}
\put( 50,60){\vector(-1,0){60}}
\put(130,65){$D$}
\put(100,60){\vector(1,0){60}}
\put( 20,65){$D$}
\put( 45,65){$a$}
\put(105,65){$b$}
\put( 75,45){$V_{\cap}$}
\put( 40,45){$V_{a}$}
\put(105,45){$V_{b}$}
\put( 75, 7){$V_{\cup}$}
\end{picture}
\caption{Union of two hyperspheres of radius $D$ centered at sites $a$ and $b$, which are a distance $D_0$ apart. There are no other sites inside the inner hyperspheres of radius $D_0$.}
\label{fig:2}
\end{figure}
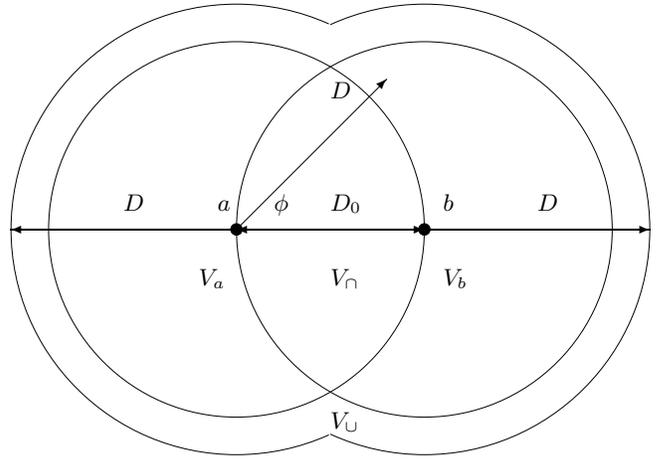
\end{center}

The joint p.d.f. that the points $a$ and $b$ form a couple and have a pair distance $D_0$ corresponds to the probability that no other points lie inside the volume $V_{\cup}(D_0,D_0)$:
\begin{eqnarray}
\label{eq:prob_d0}
P_d(D_0,a \leftrightarrow b)             &=& \frac{\mbox{d} V (D_0)}{\mbox{d} D_0}
e^{-V_{\cup}(D_0,D_0)} \\
\nonumber
        &=& A_d d D_0^{d-1} e^{- [1 + p_d(1)] V(D_0)}  \;.
\end{eqnarray}
For example, in $d=1$ we have $P_1(D_0,a \leftrightarrow b)= 2 e^{-3 D_0}$.
Notice that $P_d(D_0,a \leftrightarrow b)$ is not normalized:  integrating with respect to $D_0$ one recovers the couple density $P_d(a \leftrightarrow b)$ given by Eq.~(\ref{eq:couple_density}).

\subsection{Couple nearest neighbor distance distribution function}
\label{sec:cnnddf}

Define $D_j$ as the distance of the $j^{\mbox{th}}$ nearest neighbor ($1 \leq j \leq n$) {\em to the couple} $a \leftrightarrow b$.
For instance, the distance of the {\it first} nearest neighbor to the couple is $D_1 = \min \{D_{a1},D_{b1}\}$, that is, the minimal one between the distance from sites $a$ and $b$ to their second nearest neighbors.
Thus, the distance of the {\it second} nearest neighbor {\em to the couple} will
be the minimal one among the remaining distances ($\max \{D_{a1},D_{b1}\}$)
and $\{D_{a2},D_{b2}\}$: $D_2 = \min\{\max \{D_{a1},D_{b1}\},
D_{a2},D_{b2}\}$.

To calculate the couple's joint p.d.f. $P(D_0,D_1)$,
one asks first for the probability that the couple nearest neighbor is at a distance $r$ {\it larger} than  $D_1$.
That is, one asks for the probability that two $d$-spheres of radius
$D_1$, one centered at site $a$ and the other centered at site $b$ are empty, {\it given} the fact that the $d$-spheres of radius $D_0$
centered at the same sites are also empty.
For a Poisson process, this probability is:
\begin{equation}
\label{eq:rgtd1}
P_d(r>D_1|D_0) =  \frac{e^{-V_{\cup}(D_1,D_0)}}{e^{-V_{\cup}(D_0,D_0)}} \;     .
\end{equation}

Then, the probability density function that $r > D_1$ is:

\begin{eqnarray}
P_d(r>D_1, D_0) & = & P_d(r>D_1|D_0) P_d(D_0,a \leftrightarrow b) \nonumber \\
                & = & \frac{\mbox{d} V(D_0)}{\mbox{d} D_0}  e^{-V_{\cup}(D_1,D_0)}  \; ,
\end{eqnarray}

\noindent where $P_d(D_0,a \leftrightarrow b)$ has been given by Eq.~(\ref{eq:prob_d0}).
Taking the derivative of the above equation with respect to $D_1$, one finally obtains the probability
distribution function to have a couple $a \leftrightarrow b$ of size $D_0$ with the
first neighbor to the couple at a distance $D_1$:

\begin{eqnarray}
\nonumber
P_d(D_1, D_0,a \leftrightarrow b) & =&   \exp[-V_{\cup}(D_1,D_0)]
\frac{\mbox{d} V(D_0)}{\mbox{d} D_0}   \\
 &   &  \times \frac{\partial V_{\cup}(D_1,D_0)}{\partial D_1} \Theta(D_1-D_0) \;
\label{eq:D0D1}
\end{eqnarray}

\noindent where the Heaviside step function $\Theta(D_1-D_0)$ has been introduced to ensure that $D_1 > D_0$ and

\begin{eqnarray}
\nonumber
\frac{\partial V_{\cup}(D_1,D_0)}{\partial D_1}   &=&    \frac{V_{\cup}(D_1,D_0) d }{D_1}  \\ & & +\frac{\partial p_d(D_0/D_1)}{\partial D_1}  V(D_1)   \label{eq:der_Vu}
\end{eqnarray}

\noindent where

\begin{equation}
\frac{\partial p_d(D_0/D_1)}{\partial D_1}  =
\frac{- D_0 \left[ 1 - \left( \frac{D_0}{2 D_1} \right)^2 \right]^{\frac{d-1}{2}}}
{D_1^2 B \left(\frac{1}{2},\frac{d+1}{2} \right)} \:.
\end{equation}

\noindent For the expression of $\partial V_{\cup}(D_1,D_0)/\partial D_1$ for some values of $d$, see table \ref{tab:1}.

\subsection{Generalizing the joint distribution function}

Generalizing this procedure, the joint p.d.f. of distances for the $k^{\mbox{th}}$
nearest neighbors of the $a \leftrightarrow b$ couple is obtained:

\begin{eqnarray}
\nonumber
& & P_d(D_k, D_{k-1},...,D_0,a \leftrightarrow b) = \exp\left[-V_{\cup}(D_k,D_0)\right] \\ & & \times \frac{\mbox{d}  V(D_0)}{\mbox{d} D_0}  \prod_{j=1}^k
\left[ \frac{\partial V_{\cup} (D_j,D_0)}{\partial D_j} \Theta(D_j-D_{j-1})  \right]    \; .
\label{eq:gen_joint_dist}
\end{eqnarray}

Notice that the above result is not normalized (stressed  by the label $a \leftrightarrow b$).
To normalize it one must divide by  $P_d(a \leftrightarrow b)  = 1/[1 + p_d(1)]$ (see Eq.~(\ref{eq:couple_density})).
For instance, for $d = 1$ one has:
\begin{equation}
P_1(D_k, D_{k-1},...,D_0) =  2^k 3 e^{-(D_0 + 2 D_k)} \prod_{i = 1}^{k} \Theta(D_i - D_{i-1}) \; ,
\label{eq:jointd1}
\end{equation}
Notice that normalized p.d.f.'s do not carry the  label $a \leftrightarrow b$.

\subsection{Stability distribution function}

Let us define the ``stability'' of an arbitrary pair of points $(a,b)$ (not necessarily reflexive neighbors) to be:
\begin{equation}
\label{eq:stability}
\Delta = D_{nn} - D_{ab} \; ,
\end{equation}
where $D_{nn} =\min\{D_{ak},D_{bk}\}$ is the distance to the nearest neighbor $k$ of $a$ or $b$ not including $a$ or $b$.
Notice that if (and only if) the stability $\Delta$ is positive then $a$ and $b$
are reflexive neighbors (a couple).
This is an alternative way of defining couples and
we may write $\Delta = D_1 - D_0$. The couple stability p.d.f. is:
\begin{equation}
P_d(\Delta) =  \int_0^{\infty} \mbox{d}D_0
\mbox{d}D_1 \; P_d(D_1,D_0) \delta[\Delta - (D_1 - D_0)] \; .
\end{equation}

For example, for $d=1$ this gives:
\begin{equation}
\label{eq:pdf_stability}
P_1(\Delta) = 2 \exp (-2 \Delta) \Theta(\Delta) \:.
\end{equation}
This quantity will be used when calculating the average residence time  next section.

\section{Glass transition}
\label{sec:phase_trans}

So far we have considered the topological aspects of the network where
the walk is performed.
Now we turn our attention to the dynamics of the hopping process.
As mentioned before, for $T = 0$ the walk is deterministic.
After a transient the walker falls into a 2-cycle defined by a couple.
However, for $T>0$ the couples are not absolute traps any longer, being characterized by a residence or trapping time $t_r$.
To calculate the average residence time let $a$ and $b$ be the sites of a couple, $D_0 = D_{ab}$ their pair distance and $D_{aj}$($D_{bj}$) the distance from $a$($b$) to its $(j+1)$-th nearest neighbor. Consider first a linear cost function $E(D) = D$ (other cost functions will be discussed later).
The probabilities of transition from site $a$ to site $b$ and vice versa
are:
\begin{eqnarray}
\nonumber
p_a &\equiv& W_{a \rightarrow b}(\beta) = \frac{e^{-\beta D_0} }{ Z_a(\beta)} \; , \; \; Z_a(\beta)  =   \sum_{i=0}^{n-1} e^{-\beta D_{ai}}\label{eq:Wab} \; , \\
p_b &\equiv& W_{b \rightarrow a}(\beta) = \frac{e^{-\beta D_0} }{ Z_b(\beta)} \; , \; \; Z_b(\beta)  =   \sum_{i=0}^{n-1} e^{-\beta D_{bi}}  \label{eq:Wba} \; .
\end{eqnarray}

Suppose that at $t = 0$ the walker is at site $a$.
The probability that the walker remains inside the couple for $t$ steps and then leaves it, is (as schemed in Table
\ref{table:3}): $P_e(t) = (p_a p_b)^{ t/2} q_a$, for even
values of $t$ and $P_o(t) = (p_a p_b)^{(t-1)/2} p_a q_b$  for odd
values of $t$, with $q_a = 1- p_a$ and $q_b = 1-p_b$.
If at $t = 0$ the walker is at site $b$, one simply has to
exchange the indexes $a$ and $b$ in $P_o(t)$ and $P_e(t)$.
Since the travel may start either at site $a$ or site $b$ the probabilities that the walker remains in the cycle up to time $t$ are:
$P_e(t)  =  (p_a p_b)^{ t/2} ( q_a + q_b )/2$ and $P_o(t)  =  (p_a p_b)^{(t-1)/2} (p_a q_b + p_b q_a)/2$,
with $P(t)$ properly normalized:
$\sum_{t =0}^{\infty} P(t) = \sum_{k = 0}^{\infty} [P_e(2 k) + P_o(2 k + 1)] = 1$.

\begin{table}
\begin{tabular}{cc}
\hline
Time Step ($t$)  &   $P(t)$    \\
\hline
0             & $q_a$ \\
1             & $p_a q_b$ \\
2             & $p_a p_b q_a$ \\
3             & $p_a p_b p_a q_b$ \\
4             & $(p_a p_b)^2 q_a$ \\
5             &  $(p_a p_b)^2 p_a q_b$\\
6             & $(p_a p_b)^3 q_a$ \\
\vdots        & \vdots \\
$t$ odd       &$(p_a p_b)^{(t-1)/2} p_a q_b$ \\
$t$ even      & $(p_a p_b)^{t/2} q_a$ \\
\hline
\end{tabular}
\caption{Probability $P(t)$ that a walker, initially placed in point $a$, makes its first exit from the couple at time $t$.}
\label{table:3}
\end{table}

The expected residence time for this couple is:
\begin{eqnarray}
\nonumber
t_r & = & \sum_{t = 0}^{\infty}P(t)\:t
      =   \sum_{k = 0}^{\infty} [ P_e(2 k)\:2 k   +   P_o(2 k + 1)\:(2 k + 1) ] \\
    & = & \frac{p_a p_b + (p_a + p_b)/2}{1 - p_a p_b} \: .
\label{eq:ts}
\end{eqnarray}

The calculation of the average $\langle t_r \rangle$
(over all possible realizations of the points)
is somewhat involved, for general values of $n$.
The problem lies in the fact that the set of nearest neighbors to
site $a$ and the corresponding one for site $b$ are neither uncorrelated
(independent) nor completely correlated.

The consideration of all the possible correlations involves a
number of integrals that increases exponentially with $n$.
Nevertheless it is possible to calculate bounds for $\langle t_r\rangle$ which reveal its characteristic diverging behavior.
To obtain these bounds, we next present a simple approximation which is physically intuitive followed by improvements to this calculation.

\subsection{Symmetrical and asymmetrical approximation}
In the symmetrical approximation, for $n=2$, the transition probabilities corresponding to the two members of a couple are taken to be equal: $p_a = p_b = p = e^{-\beta D_0}/Z $, using $Z_a = Z_b = Z = e^{-\beta D_0} + e^{-\beta D_1}$, with $D_1 = \min(D_{1a},D_{1b})$  (as defined in Sec.~\ref{sec:cnnddf}). The residence time is then given by  $t_{rs} = p/(1 - p) = e^{\beta \Delta}$, where $\Delta = D_1 - D_0$ is the stability of the couple, defined by Eq.~(\ref{eq:stability}). This would be exact if the nearest neighbors to the couple were degenerated, that is, if there was one point at a distance $D_1$ from $a$ and a point at a distance $D_1$ from $b$.

In the general case, $t_{rs}$ is a lower bound to $t_{r}$. This stems from the very definition of $D_1$: for one of the points of the couple the distance to its second nearest neighbor (SNN) is, in the general case, necessarily {\it larger} than $D_1$. Thus, given $D_1$, a couple whose members have their respective SNNs at the same distance, $D_1$, has the shortest possible residence time.

In the asymmetrical approximation, for the member of the couple (say, $b$) whose SNN is at a distance larger than $D_1$, the transition probability to the SNN is taken to be zero, that is $p_b = 1$. This leads to a residence time $t_{ra} = 2 e^{\beta \Delta} + 1/2$. This residence time clearly bounds the true residence time from above, because we are assuming that the walker cannot leave the couple from point $b$.

From these approximations, one obtains:
\begin{equation}
e^{\beta \Delta} \le t_r \le 2 e^{\beta \Delta} + \frac{1}{2} \; .
\end{equation}

The average over the disorder is calculated using Eq.~(\ref{eq:pdf_stability}):
\begin{eqnarray}
\nonumber
\langle e^{\beta \Delta} \rangle & = & \int_0^{\infty} \mbox{d} \Delta\: P_1(\Delta)\:
e^{\beta \Delta}
                                   =  2 \int_0^{\infty} \mbox{d}\Delta
                                   e^{-\Delta(2 - \beta)}  \\
     & = & \frac{2}{2 - \beta} \; ,
\end{eqnarray}
for $\beta < \beta_c = 2$.

Thus, the average residence time is bounded as:
\begin{equation}
\label{eq:bounds1}
2 \le (2 - \beta) \langle t_r \rangle \le  5 - \frac{\beta}{2}
\end{equation}
and diverges when $\beta \rightarrow \beta_c = 2$.
These bounds are shown in Fig.~\ref{fig:3}.

Above the critical temperature the walker falls into traps (2-cycles) which have finite average residence time.
This allows normal diffusion for long time scales.
However the walker eventually falls into a sink which is analogous to an absorbing state.
Notice that in general sinks are not couples.
Sinks are absolute traps arising from the finite connectivity $n$.
Couples are dynamical traps and temperature allows the escaping from couples but not from sinks.
Below the critical temperature $T_1 = 1/\beta_1 = 1/2$  the system
falls into an out-of-equilibrium ``glassy'' phase where the walker explores
deeper and deeper (more stable) couples and presents aging phenomena like in
Bouchaud's trap model \cite{bouchaud:1992}.
This scenario has been called ``weak ergodicity breaking'' by Bouchaud and
co-workers \cite{bouchaud:2001}.

\subsection{Improved bounds for the residence time}
\label{sec:app1}

The previous simple calculations give insight on the physical mechanism
involved in the divergence of the residence time, but we can
obtain better estimates for the bounds on the residence time, which work for general values of $n$.
Using the ordered set of distances, we have that $D_j \le D_{a(b)j} \le D_j+D_0$.
The second inequality expresses the fact that our definition of the
set of values $D_j$ implies that a $d$-sphere of radius $D_j+D_0$,
centered in point $a$ (or $b$) contains at least $j$ points.
Thus, we can bound the normalization factors in Eq.~(\ref{eq:Wab})
as follows: $Z_{2} \leq  Z_{a(b)} \leq Z_{1}$, where $
Z_{1} =  e^{-\beta D_0} + \sum_{j=1}^{n-1} e^{-\beta D_j}$ and $Z_{2}  =  e^{-\beta D_0}[1  + \sum_{j=1}^{n-1} e^{-\beta D_j}]$.
This implies that $p_1 \leq p_{a(b)} \leq p_2$, with
$p_{1}=e^{-\beta D_0}/Z_1$ and $p_{2} = e^{-\beta D_0}/Z_2$,
which allows us to bound the residence time:
$t_{r_1} \leq t_r \leq t_{r_2}$, with $t_{r_1} = p_1/(1-p_1)$ and $t_{r_2} =  p_2/(1-p_2)$.

To average over the disorder, one needs the joint p.d.f. of the distances of the nearest neighbors to the couple (Eq.~(\ref{eq:gen_joint_dist})).
For $d=1$ one obtains, again for $\beta < \beta_c = 2$:
\begin{equation}
\label{eq:bounds2}
2  \leq   (2-\beta) \langle t_r \rangle \leq \frac{6}{3-\beta}
\;\;\;\; \mbox{for} \;\;\;\; n=2 \; ,
\end{equation}
\noindent and
\begin{equation}
I_{\beta}(n) \leq (2-\beta) \langle t_r \rangle
\leq \frac{3 I_{\beta}(n)}{3-\beta} \;\;\;\; \mbox{for} \;\;\;\; n>2\:, 
\end{equation}

\noindent with

\begin{eqnarray}
I_{\beta}(n) & = & 2 \int_0^{\infty} \frac{\prod_{i=1}^{n-2}
(\mbox{d}x_i \, e^{-x_i})}{1+\sum_{j=1}^{n-2} \prod_{i=1}^{j} e^{-\beta x_i/2}} \; \nonumber\\
& = & \int_0^1 \frac{\prod_{i=1}^{n-2}
\mbox{d}y_i \,}{1+\sum_{j=1}^{n-2} \prod_{i=1}^{j} y_i^{\beta/2}}.
\end{eqnarray}

\begin{figure}[htb]
\begin{center}
\includegraphics[width=8.5cm]{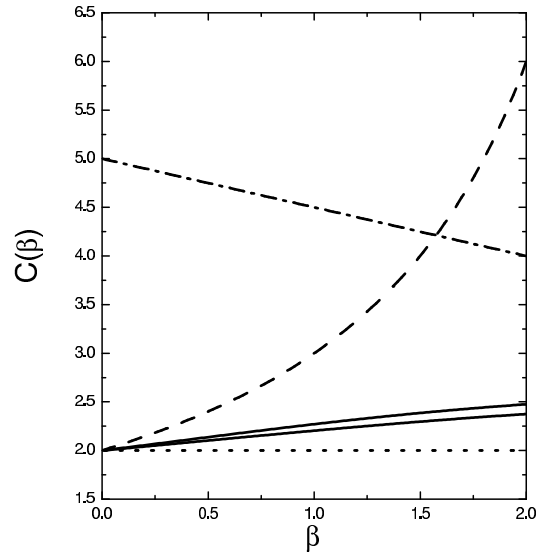}
\end{center}
\vspace{-6cm}
\caption{Bounds for the average residence time for $d=1$ with $n = 2$.
Here $C_{-}(\beta) \le (2 - \beta) \langle t_r \rangle \le C_{+}(\beta)$.
 The ascending dashed curve is the upper bound given by Eq.~(\ref{eq:bounds2}) and the descending line is the upper bound given by Eq.~(\ref{eq:bounds1}). The horizontal line is the lower bound $C_- = 2$ common to both equations. Solid lines are the tighter $C_\pm$ bounds obtained in Appendix \ref{sec:appx}.}
\label{fig:3}
\end{figure}

In Fig. \ref{fig:3} we give an example of the relationship between these bounds
and the exact value of $\langle t_r \rangle$, for the case $n=2$. In fact, the comparison is made not with the exact value but with very tight bounds to it (solid lines in Fig.~\ref{fig:3}) that have no simple analytical form and have only been obtained for the particular case of $d=1$ and $n=2$ (see Appendix~\ref{sec:appx}).

It is easy to see that $2/(n-1)<I_{\beta}(n)<2$ for all values of $\beta$.
Thus, by looking at the bounds we can deduce that $\langle t_r \rangle$ too diverges when
$\beta \rightarrow \beta_c=2$, for all finite values of $n$.

\begin{figure}[htb]
\begin{center}
\includegraphics[width=8.5cm]{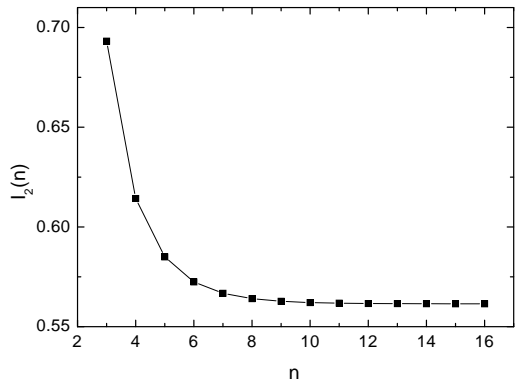}
\end{center}
\vspace{-1cm}
\caption{Numerical calculation of $I_2(n)$ as a function of $n$ showing that $I_2(n)$ tends to a positive constant.}
\label{fig:4}
\end{figure}

The function $I_2(n)$ can be proven to be a decreasing function of $n$. But it would be interesting to know whether it tends to a positive constant as $n \rightarrow \infty$, or it tends to zero.
We have not been able to calculate this limit analytically.
Nevertheless numerical results (see Fig. \ref{fig:4})
show that $I_2(n)$ seems to decrease toward a positive constant ($\sim 0.56$).
This suggests that a glass transition also exists in the limit $n \rightarrow \infty$.

Moreover, as $n$ grows, the decay of $I_2(n)$ is very slow compared to the exponential decrease of density of sinks ($C(n) \sim 4/3^n$).
Thus, the divergence of the residence time should not depend on couples inside the sinks; it seems to be produced only by couples in the GSCC. This suggests that the glass transition occurs to walkers in the GSCC before they fall into sinks.

\subsection{Distribution of residence times}

The origin of the glass transition that appears in $d=1$ is related to the trapping times p.d.f. $P(t_r)$.
Exact analytical results are difficult to obtain.
Nevertheless the probability distribution for the approximate residence times $t_{r_1}$ and $t_{r_2}$ (defined in the preceding section) gives an idea of its behavior.
For the case $n=2$, we obtain:
\begin{eqnarray}
\nonumber
P_{1}(t_{r_1}) & = & \int_0^{\infty} dD_0 \, dD_1 \, P_d(D_0, D_1) \delta\left[t_{r_1} - e^{-\beta(D_0-D_1)}\right]  \\
                   & = & \frac{2}{\beta} t_{r_1}^{-(2/\beta+1)} \\
\nonumber
P_{2}(t_{r_2}) & = & \int_0^{\infty} dD_0 \, dD_1 \, P(D_0, D_1) \delta\left[t_{r_2}
- e^{\beta(D_1)}\right] \\
                   & = & \frac{6}{\beta} \; t_{r_2}^{-(2/\beta+1)} \; \left[ 1-
                   \frac{1}{t_{r_2}^{\beta}} \right] \; .
\end{eqnarray}
The divergence of the average residence time is a direct consequence of the fact that the tail of these distributions decays with exponent $\gamma < 2$ when $\beta>2$.
Thus, the behavior of the walker is dominated by couples which have far apart nearest neighbors.

\section{Generalization to higher dimensions and different cost functions}

For higher-dimensional systems ($d > 1$), the calculations of the
properties of the system become harder but some estimates
can be obtained.

\subsubsection{Asymptotic behavior of the residence times}

It is possible to calculate the behavior of $\langle t_{r_1} \rangle$ and $\langle t_{r_2} \rangle$ when $n=2$.
For $d \geq 2$ and $\beta \gg 1$, the calculated asymptotic bounds to $\langle t_{r} \rangle$ are:
\begin{equation}
\frac{1}{\left[ d+B\left(\frac{d+1}{2},\frac{1}{2}
\right)\right]^d} \leq
\frac{\langle t_{r} \rangle}{C_d \exp[{E_d \beta^{d/(d-1)}}]}   \leq
1
\end{equation}
with
\begin{eqnarray}
E_d & = & (d-1)({d A_d^{1/d}})^{-d/(d-1)} \; , \\
C_d & = &\sqrt{\frac{2}{d-1}} \,  B\left(\frac{d+1}{2},\frac{1}{2}
\right)^d \, \Gamma \left(\frac{d}{2}+1 \right)^\frac{3-2d}{2d-2} \nonumber \\
    &   &\times d! \, d^\frac{2d^2-3}{2d-2} \, \pi^\frac{d^2-d/2-1}{2d-2} \; ,
\end{eqnarray}
where $A_d$ is given by Eq.~(\ref{eq:ad}).

Thus, there is no glass transition when $d \geq 2$ for the linear cost function $E(D_{ij}) = D_{ij}$.
However, as we show below, there is a specific cost function which yields a glass transition for each dimension.

\subsubsection{Generalized cost functions}

Consider the general case where the transition rates have the form
$W_{i\rightarrow j} = \exp[-\beta E(D_{ij})]/Z_i$ with an arbitrary cost
function $E(D_{ij})$.
In the symmetric approximation, the average residence time is:
\begin{equation}
\langle  t_{rs} \rangle  = \int_0^{\infty} \mbox{d}D_1 \mbox{d}D_0
P_d(D_1,D_0) e^{\beta [E(D_1)-E(D_0)]}  \; ,
\end{equation}
where $P_d(D_1,D_0)$ is given by Eq.~(\ref{eq:jointd1})(using $k=1$).
It is clear that the two competing factors which create the possibility of a glass
transition are: {\em i)} the tail of $P_d(D_1,D_0)$ and {\em ii)} the exploding $\exp[\beta E(D_1)]$
factor.
As it can be seen from Eq.~(\ref{eq:gen_joint_dist}),
the tail of $P_d(D_1,D_0)$ is: $P_d(D_1,D_0) \propto \exp(- A_d D_1^d)$, where $A_d$ is given by Eq.~(\ref{eq:ad}).
Thus, the symmetric approximation leads to:
\begin{equation}
\label{eq:alfa}
\langle  t_{rs} \rangle  \propto \int_{D_0}^{\infty} \mbox{d}D_1 \; e^{\beta E(D_1)- A_d D_1^d }  \; .
\end{equation}

Consider, for instance, the family of cost functions: $E = D_{ij}^\alpha$.
If $\alpha$ is smaller than $d$, the residence time is always finite and no glassy behavior is
observed (as we have previously observed in the case $d=2$ and $\alpha=1$).
The limiting case $T = 1/\beta \rightarrow 0$ will be examined elsewhere.
On the other hand, if $\alpha > d$, the system is always in the glassy state for any value of $\beta$.

If the cost function exponent is equal to the space dimensionality, $\alpha = d$, the two terms in the argument of exponential function in Eq.~(\ref{eq:alfa}) can compete independently of the value of $D_1$ and a glass transition occurs at $\beta_d = A_d = \pi^{d/2}/\Gamma(d/2 + 1)$.
For example, $\beta_1 =2$ for $d=1$ (as seen previously), $\beta_2 = \pi $, for $d=2$ and $\beta_3 = 4\pi/3$, for $d=3$.

This kind of glass transition generalizes those found in trap models.
In trap models, once the tail of the barrier distribution is given (say, a gaussian tail as in Ref. \onlinecite{bouchaud:2002}), one cannot carpenter the Arrhenius term to compete with it.
This is different in our exploratory random walk model.
In our case what is given is not the barrier distribution $P(E)$,
but the distance distribution $P(D_{ij})$ which characterizes the landscape.
Since the transition probability for exploratory random walks
depends on an arbitrary cost function $E(D_{ij})$, one can always find what condition
must be satisfied by the cost function so that a glass transition occurs.

\subsubsection{L\'evy flights}

From Eq.~(\ref{eq:alfa}) one sees that any $\alpha < d$ prevents the glass transition.
In particular, one can study the limit $\alpha \rightarrow 0$, i.e., $E(D_{ij}) = \ln D_{ij}$.
In this case, a hopping process with power law tail, $W_{i\rightarrow j } = D_{ij}^{-\beta}/\sum_{j\in nn} D_{ij}^{- \beta}$ is found.
Transition probabilities with power law (``Levy'') tails lead to finite average residence
times for any value of $d$.
Such L\'evy exploration process has been considered, for instance, in  Refs. \onlinecite{viswanathan1,barkai:1998,viswanathan2}.

But glass transitions may occur if the distribution of point distances also has power
law tails. For example, suppose that the points coordinates are drawn from some
distribution that produces $P(D_1,D_0) \propto D_1^{-b}$. Then, the diverging part of the
residence time integral has the form
$ \langle  t_{rs} \rangle  \propto \int \mbox{d}D_1  D_1^{\beta-b}$.
This means that a glass transition occurs if $\beta \ge \beta_c = b - 1$.

\section{Conclusion}
\label{sec:conc}

We have introduced a random walk in a disordered medium dependent
on a control parameter which tunes the system from an ergodic to a
non-ergodic regime through a glass transition. The random walk is
similar to a hopping process in a multivalley landscape. The $d =
1$ ``local minima'' or traps emerge as 2-cycle attractors produced
by sites which are mutual nearest neighbors. For $d=1$ these traps
present a power law distribution of trapping times and the average
trapping time diverges for $T \le T_c =1/2$, leading to a
``glass'' transition in the weak ergodicity breaking scenario of
Bouchaud's trap model \cite{bouchaud:1992,monthus:1996}. Given any
distribution of distances and any dimensionality, one can always
find a cost function for which a similar glass transition occurs.

The analytical treatment to estimate the bounds of the average residence time has been possible because we have considered finite connectivity. The finite value of $n$ is also responsible for the appearance of special sets of points (sinks and sources) in the directed graph where the walks are performed. Sinks are absolute, inescapable topological traps insensitive to temperature.
Couples (that produce 2-cycles) are relative traps that can be escaped with finite temperature.
We have considered a possible glass transition involving couples
before the walker falls into sinks.
Given the behavior of the bounds to the average residence time, we also expect a glass transition to appear in the fully connected network with $n = N-1$, where no sinks exist, only couples.

Concerning the application of our model to exploratory behavior,
one (a posteriori) obvious conclusion is that free exploration
(diffusion or super diffusion) occurs when the tail of the hops
p.d.f. decays slower than the tail of the first neighbors
distances p.d.f. However, if costs are associated to the
travelling distance, the marginal scenario, where the two p.d.f.
tails ($\alpha = d$) compete, may be of interest. We conjecture
that working at the glass transition border may be optimal when
high costs are associated to travel distances (this will be
presented elsewhere).

The 2-cycle phase (at zero temperature) and the glassy behavior of
the stochastic Tourist Walk may be compared to similar behaviors
in neural and Kauffman networks. Such comparison is not possible
by using the Random Map. The random walk in a disordered graph
introduced here is interesting because it generalizes the usual
trap models: in the Tourist Walk traps are made of dynamical
cycles instead of energy minima. We are currently studying the
stochastic Tourist Walk with memory $\tau > 0$, so that traps are
cycles with large periods (similar to what happens in Kauffmann
networks and asymmetric neural networks~\cite{bastolla:1998}). We
expect that the relevant features should be the same as those
observed in the present work. In short, if one wants to escape
from vicious cycles, small perturbations may not be sufficient:
freedom only appears above a finite temperature $T_c$.

\begin{acknowledgments}
The authors acknowledge useful conversations with J. Arenzon, N.
Caticha, R. Dickman,  J. R. Drugowich de Felício, J. F. Fontanari,
M. A. Idiart, G. F. Lima, D. A. Stariolo, A. C. Roque da Silva, R.
da Silva and C. Yokoi. O. Kinouchi has been
supported by FAPESP.
\end{acknowledgments}

\appendix
\section{Improved bounds}
\label{sec:appx}

In this appendix we derive tighter bounds than those given by Eqs.~(\ref{eq:bounds1})
and~(\ref{eq:bounds2}), but they are only valid in the special case of $d=1$ and $n=2$.

Consider first a couple of points, $a$ and $b$, separated by a distance $D_0$.
To calculate $Z_a$ and $Z_b$ (see Eq.~(\ref{eq:Wba})) one only needs the positions of $a$ and $b$ and their two nearest neighbors.
The first neighbor to each of them is, by definition, the other member of the couple.
Since the distribution of the distances of the neighbors {\it to the couple} is known, one needs to determine which of these neighbors is the second-nearest neighbor (SNN) to $a$ and which is the SNN to $b$.

Without loss of generality we assume $b$ to be the rightmost point of the couple.
The first neighbor to the couple (called FNNC) is taken to be placed at a distance $D_1$ (with $D_1>D_0$), and at the right of $b$, what makes it the SNN to $b$.
Now the task is reduced to the determination of the SNN to point $a$.

When one considers the position of the second nearest neighbor to the couple (called SNNC), at a distance $D_2$ (with $D_2>D_1$), three possibilities arise for the SNN of $a$:

\renewcommand{\theenumi}{\Roman{enumi}}
\begin{enumerate}
 \item $D_2>D_1+D_0$. In this case, the SNN to $a$ is the FNNC, and its distance to point $a$ is: $D_1+D_0$.
 \item $D_2<D_1+D_0$, but it is placed to the {\it left} of $a$.
        In this case, the SNN to $a$ is the SNNC, and its distance to point $a$ is: $D_2$.
 \item $D_2<D_1+D_0$, but it is placed to the {\it right} of $b$.
       In this case one needs to know the position of further neighbors to the couple to obtain the SNN to $a$.
       If they are placed at distances larger than $D_1+D_0$, then the SNN to $a$ would be the SNNC.
       Otherwise, one would have to know which neighbors, of those having distances smaller than $D_1+D_0$, are placed to the left of $a$.
 \end{enumerate}

The corresponding residence times can be calculated exactly for cases I and II.

For case I, the partition functions are: $Z_{aI}(\beta) =  e^{-\beta D_0}+e^{-\beta (D_0+D_1)}$ and
$Z_{bI}(\beta) =  e^{-\beta D_0}+e^{-\beta D_1}$. For case II they are: $Z_{aII}(\beta)  =   e^{-\beta D_0}+e^{-\beta D_2}$ and $Z_{bII}(\beta)  =  e^{-\beta D_0}+e^{-\beta D_1}$.

Eqs.~(\ref{eq:ts}) and~(\ref{eq:Wba}) yield the residence times:
\begin{eqnarray}
\nonumber
t_{rI}(\beta) & = &  \frac{(1+e^{-\beta D_0})/2+2e^{\beta (D_1-D_0)}}{1+e^{-\beta D_0}+e^{-\beta D_1}}                       \\
t_{rII}(\beta) & = & \frac{2e^{-\beta D_0}+(e^{-\beta D_1}+e^{-\beta D_2})/2}{e^{-\beta D_1}+e^{-\beta D_2}+e^{\beta (D_0-D_1-D_2)}}    \; .
\end{eqnarray}

The problem with $t_{rIII}$ is that to calculate it exactly one needs to break case III into an infinity of particular cases.
For our purposes it suffices to calculate some bounds to its real value. As it has been discussed in case III, the distance of point $a$ to its SNN satisfies: $D_2 < D_{SNNa} < D_0 + D_1$.
Therefore, it is clear that the corresponding residence time must satisfy: $t_{rI} < t_{rIII} < t_{rII}$.

Remembering that the probability for a given point to be at one side of the couple is $1/2$, and averaging, one obtains the bound:
\begin{eqnarray}
&&\int dD_0 dD_1 \, dD_2 P_1(D_0,D_1,D_2) \, [t_{rI} \Theta(D_2-D_1-D_0) \nonumber \\
&&+\frac{t_{rI}+t_{rII}}{2} \; \Theta(D_1+D_0-D_2)] \nonumber  \\
&& \leq \,\, \langle t_r \rangle \,\, \leq \nonumber  \\
&&\int dD_0 \, dD_1 \, dD_2 P_1(D_0,D_1,D_2) [t_{rI} \Theta(D_2-D_1-D_0) \nonumber  \\
&&+t_{rII} \, \Theta(D_1+D_0-D_2)]  \; .
\end{eqnarray}

We have calculated these integrals numerically, and the resulting bounds are presented as full lines in Fig.~\ref{fig:3}.

\bibliographystyle{apsrev}

\end{document}